# Double Degree of Freedom Helmholtz Resonator Based Acoustic Liners

Abhishek Gautam[1], Alper Celik[2] and Mahdi Azarpeyvand[3]
*Faculty of Engineering, University of Bristol, Bristol, United Kingdom*

A numerical and experimental study on double degree of freedom Helmholtz resonator based acoustic liners is performed in this paper, with the motivation to improve the understanding of their sound attenuation mechanism. A single degree of freedom liner is designed and manufactured, as baseline, to compare with the double degree of freedom acoustic liners. The Grazing Flow Impedance Tube Facility at Bristol is used to measure the transmission coefficients of the acoustic liner samples, which are compared against data obtained via numerical simulations on Comsol. Both sets of data show good agreement. The effect of changing the internal chamber dimensions is studied numerically, through Comsol steady state simulations, followed by time domain simulations. A minimum, in the difference between the primary and secondary resonance peaks is observed. This may be an indication of the presence of an optimum volume ratio of the two internal cavities in the double degree of freedom resonator, for a broad bandwidth sound absorption.

## I. Nomenclature

| | | |
|---|---|---|
| *DoF* | = | Degree of freedom |
| *TL* | = | Transmission loss |
| *L* | = | Cavity height of baseline single degree of freedom liner |
| *L1* | = | Height of the first cavity for the double degree of freedom liner |
| *L2* | = | Height of the second cavity for the double degree of freedom liner |
| *T* | = | Neck length of the single degree of freedom liner |
| *T1* | = | Neck length of the first neck for the double degree of freedom liner |
| *T2* | = | Neck length of the second neck for the double degree of freedom liner |
| *V0* | = | Cavity volume for the baseline single degree of freedom liner |
| *V1* | = | Volume of first cavity for the double degree of freedom liner |
| *V2* | = | Volume of second cavity for the double degree of freedom liner |
| *PML* | = | Perfectly matched layer |
| $f_0$ | = | Baseline single degree of freedom liner resonance frequency |
| Ψ | = | Ratio of double degree of freedom resonator first cavity height to baseline single degree of freedom liner cavity height |

## II. Introduction

Acoustic liners are often used on the inner wall of the intake and bypass ducts of aircraft engines as the primary devices for reducing the aeroengine noise. The most common liners are based on Helmholtz resonators, where the sound field in the main duct communicates with a cavity through a neck. However, Helmholtz resonators have an inherited disadvantage of affecting a narrow bandwidth of sound spectra. Development of new, more efficient aero engines has led to the requirement of acoustic liners having broadband noise absorption. Industrial multi- degree of freedom liners are liners with a double-layer sandwich construction where the two layers of honeycomb are separated by a porous septum. Having a porous septum between the two layers improves the sound absorption characteristics of the whole liner and the septum can be designed to have noise suppression characteristics over a large range of frequencies. Multiple changes to the geometry have been studied in the past to improve the noise suppression characteristics of Helmholtz resonators [6-11].

---

[1] PhD. student, Mechanical Engineering, abhishek.gautam@bristol.ac.uk.
[2] Post-Doctoral Research Associate, Mechanical Engineering, alper.celik@bristol.ac.uk.
[3] Professor of Aerodynamics and Aeroacoustics, Mechanical Engineering, mahdi.azarpeyvand@bristol.ac.uk.

The effect of changing the neck shape, neck length and having a perforated extended neck, on the sound transmission loss was studied by Selamet and Lee [1]. Xu et al. [2] established the acoustics characteristics of a two degree of freedom Helmholtz resonator in terms of the lumped parameter theory. Helmholtz resonators having tapered necks with the area of the neck expanding towards the cavity was analysed both theoretically and experimentally by Tang [3]. A one-dimensional duct with a coupled resonator mounted on the side wall was analytically developed by Griffin et al. [4]. A lumped analysis was used by Wan and Soedel [5] to obtain expressions for the resonance frequencies of two degree of freedom helmholtz resonators.

Although there have been multiple studies to create analytical models for single and multi-degree of freedom helmholtz resonators, the focus of the current research is to improve the understanding of the underlying mechanism of sound attenuation in a multi degree of freedom liner by:

(1) Understanding how the two resonance frequencies of a dual helmholtz resonator change with changing internal volume ratios when compared to a baseline single degree of freedom resonator.

(2) Transient analysis of the changing volume ratios for the dual helmholtz resonators to further understand the sound absorbing mechanisms in the resonators.

### III. Experimental Setup

Experiments are carried out at the University of Bristol Grazing Flow Impedance Tube Facility. The schematic of the facility is presented in Figure 1. The Impedance Tube is a 50.4mm x 50.4mm square cross-section tube, 4000 mm in length with a 3000 mm long diffusing section to reduce air velocity and minimise acoustic reflections back into the test section. Flow speeds of up to Mach 0.3 can be achieved within the test section using a 15kWh centrifugal fan. Sound pressure levels of up to 130dB were achieved in the test section via two BMS 4592ND compression drivers and microphone data was obtained via GRAS 40PL microphones. Data acquisition was achieved with a National Instruments PXIe-1082 data acquisition system. MatlabR2016a was used to interface between the data acquisition system and the signal generator to run the data acquisition code.

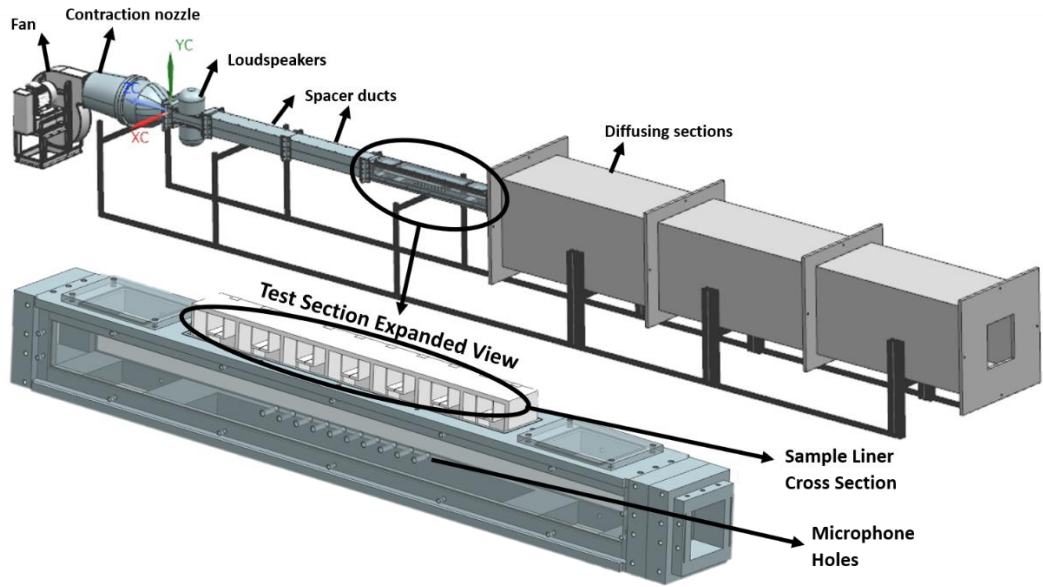

**Figure 1. Grazing Flow Impedance Tube at the University of Bristol**

For experimental analysis, a simple one degree of freedom resonator tuned to 670Hz ($f_0$) was additive manufactured alongside nine double degree of freedom liner cases where the volumes of the two chambers were a percentage of the volume V0 of the baseline single degree of freedom liner. Schematics of the two types of liner cavities can be seen in Figure 2.

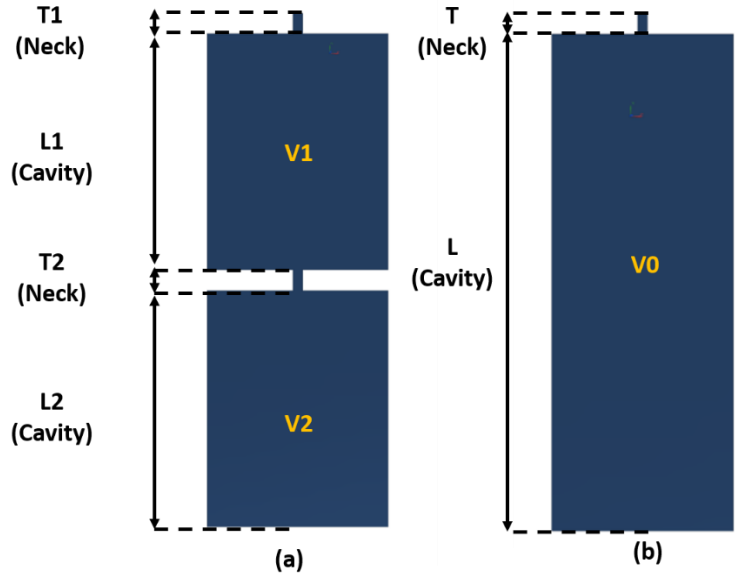

**Figure 2. (a) 2 DoF Liner cavity schematics; (b) Baseline Single Degree of Freedom Liner Schematics**

V1 and V2 are the volumes of the two internal cavities of the double degree of freedom Helmholtz resonator. T1 and T2 correspond to the thickness of the two necks. The resonator is flush mounted to the side wall of the Impedance Tube such that Neck with thickness T1 sees the incoming sound field. The cavity volumes for the double degree of freedom liner options were altered by changing the parameter $\Psi$. $\Psi$ is defined as the ratio of L1/L and is swept from 0.2 – 0.9 for the experimental study. Since the study is two dimensional and the width of each chamber is kept constant throughout the study, the length ratio governs the change in volume for the chambers. A $\Psi$ of 0.2 means that the height, L1, for Cavity 1 is 20 percent of the height, L, for the baseline single degree of freedom liner. Figure 3 shows an additive manufactured double degree of freedom liner sample, with a $\Psi$ of 0.3. Figure 4 shows CAD geometry for a selection of double degree of freedom liners designed for this study.

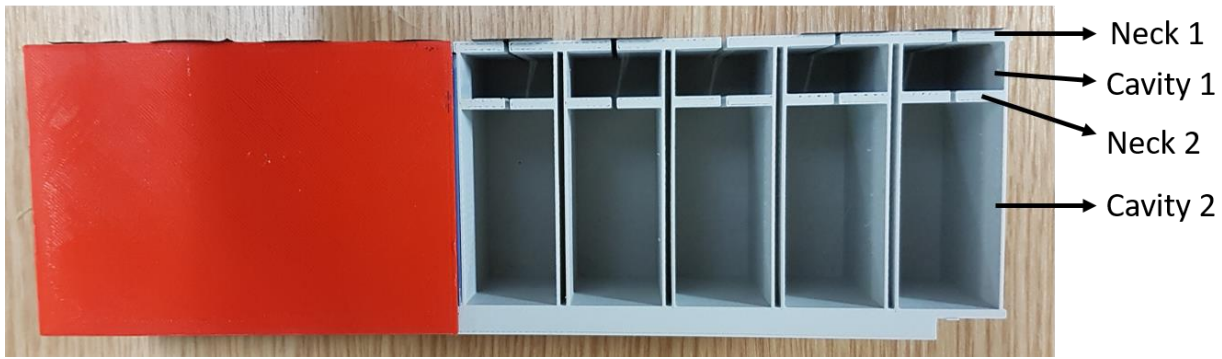

**Figure 3. Additive manufactured double degree of freedom liner option**

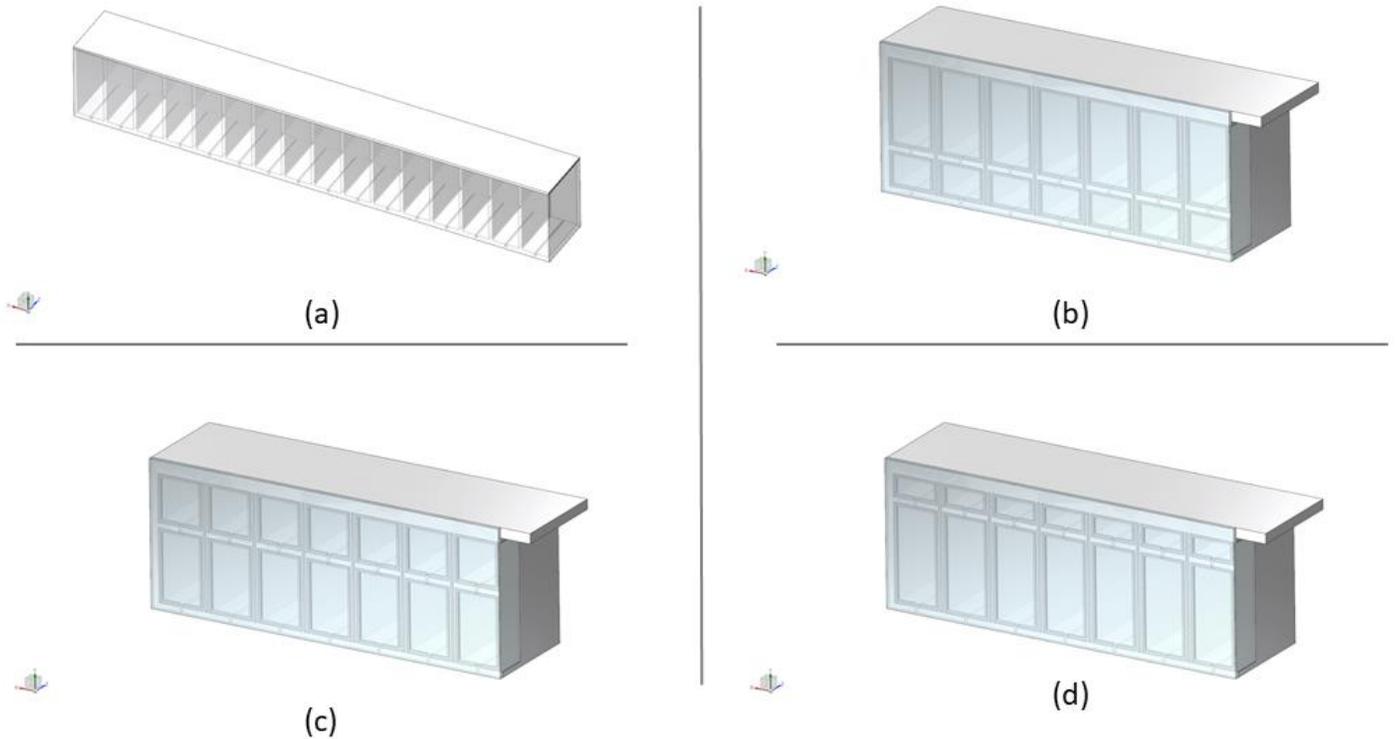

**Figure 4. CAD models for a selection of the liner concepts tested numerically and experimentally. (a) Baseline single degree of freedom liner; (b) 2DoF liner ($\Psi = 0.3$); (c) 2DoF liner ($\Psi = 0.6$); (d) 2DoF liner ($\Psi = 0.8$)**

## IV. Numerical Setup

Numerical simulations were carried out on COMSOL Multiphysics to obtain the transmission loss for the acoustic liners. A free triangular mesh was used for the impedance tube domain and a mapped mesh type was used for the acoustic box and acoustic termination. A Perfectly Matched Layer (PML) was used on either side of the impedance tube to prevent any reflections from affecting the results. The maximum element size for the mesh was chosen to be 2mm which is smaller than 1/6th of the wavelength of the highest frequency of interest i.e. 3000Hz. The minimum element size chosen was 0.12mm. The maximum element growth rate was 1.2 with a Curvature factor of 0.3 and Resolution of narrow regions of 3. The mesh quality can be seen in Figure 5. A background pressure field is applied to create a travelling plane wave. Transmission loss peaks obtained from steady state simulations were used to guide the transient simulations which were only computed at the peak frequencies to save computational time.

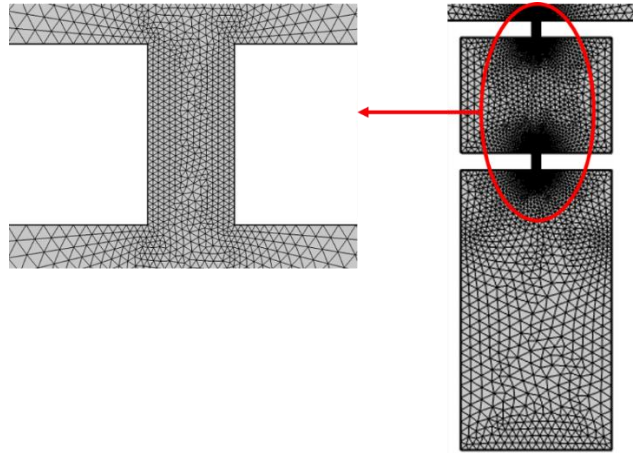

**Figure 5. Free triangular mesh used for comsol simulations**

## V. Results

A single degree of freedom liner with 16 chambers was additive manufactured and the transmission loss induced by it was analysed in the Grazing Flow Impedance Tube. Figure 6 shows the experimental results of the transmission loss induced by both the 16 chamber and single chamber liner configurations. These experimental results are compared to numerical results obtained by simulating similar geometries on Comsol. The experimental and numerical results agree well for the 16-chamber liner configuration. There is a slight deviation in the transmission loss peak between experimental and numerical results for the single chamber configuration. This can be attributed to the fact that there is a finite input power to the speakers. The Comsol simulations assume a perfect sound hard boundary whereas in the case of experiments, the liners being additive manufactured using PLA, might not be completely sound hard which may be another cause for the deviation.

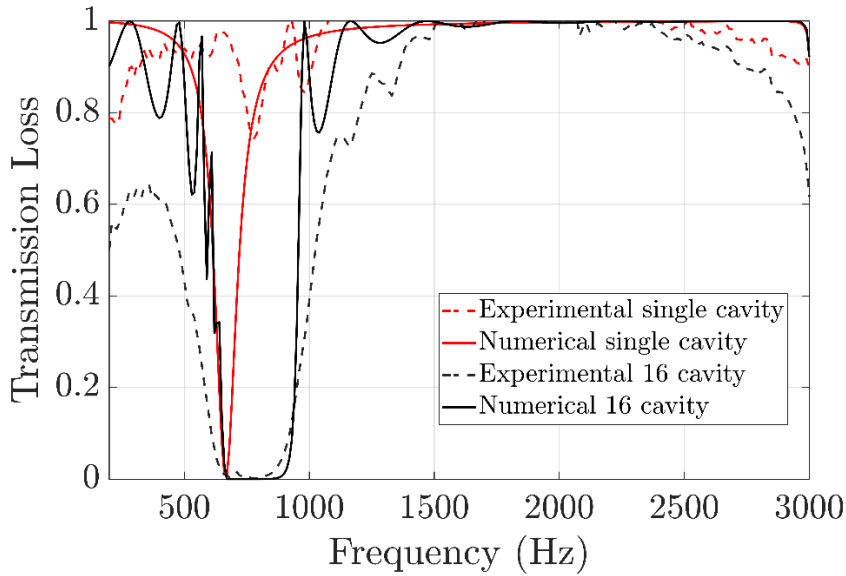

**Figure 6. Comparison of experimental and numerical simulation results for the Baseline single degree of freedom liner in both single and 16 chamber configurations**

Similar studies were performed with double degree of freedom liners. The results show a double degree of freedom liner exhibiting a broadband sound absorption behavior due to the presence of a secondary absorption peak as compared to the baseline single degree of freedom liner. The volume ratio of the chambers in the 2DoF liners was changed by altering Ψ. Figure 7 shows experimental and numerical transmission loss results for three different Ψ configurations.

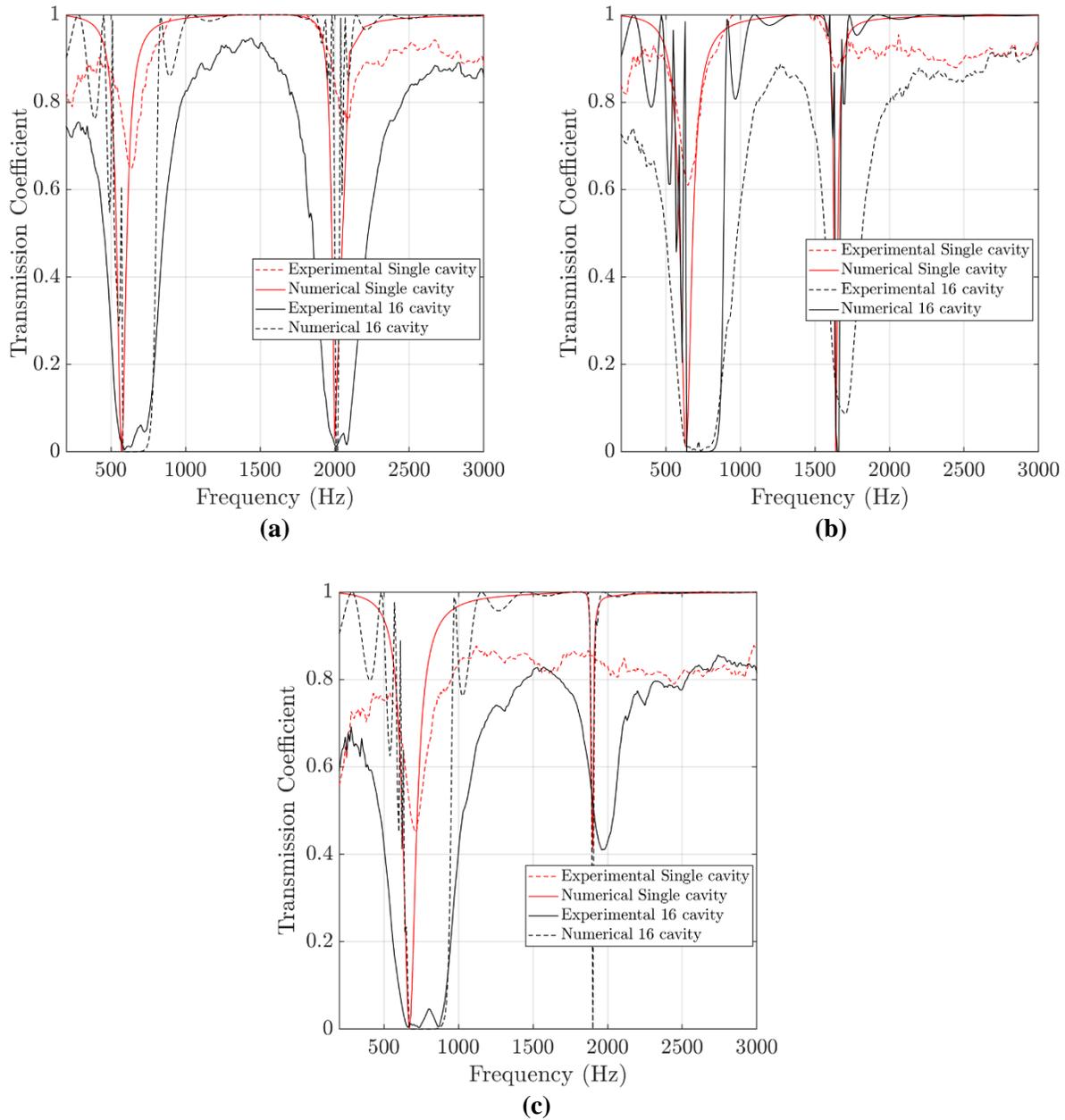

**Figure 7. Experimental and numerical simulation results for the double degree of freedom liner options in both single and 16 chamber configurations. (a) Ψ = 0.3; (b) Ψ = 0.6; (c) Ψ = 0.8**

Normalised to the baseline single degree of freedom liner resonance frequency, the primary transmission loss peak for 2 DoF liners does not show much deviation however the secondary peak is sensitive to changing volume ratio. As the volume of the first cavity increases, the difference between the first and second peak frequencies reduces up to a minimum after which it increases. Transient simulations were carried out to further investigate this convergent-divergent behaviour of the second peak frequency. 135 domain probe points were used along the length of a 2DoF liner cavity, to investigate the change in total acoustic pressure inside the cavity compared to the duct, in the time domain. Figure 8 shows how the primary and secondary peak frequencies differ with changing values of Ψ for the 2DoF liner options as compared to a baseline single degree of freedom liner.

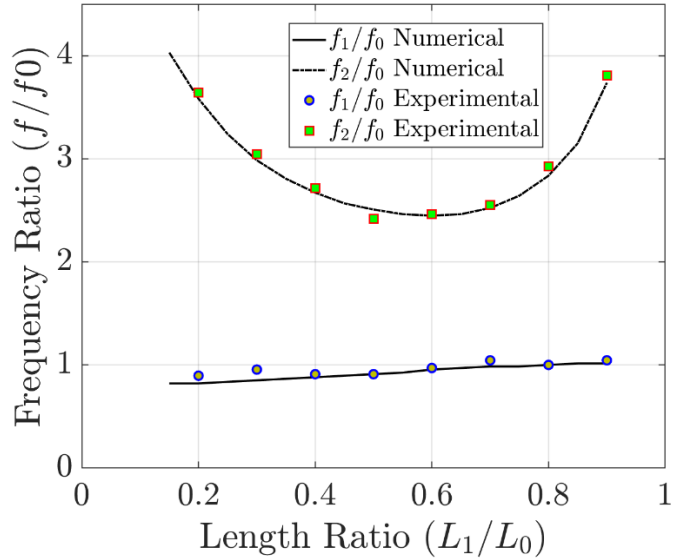

**Figure 8. Frequency ratio comparison with respect to changing length ratio (Ψ)**

Contour plots of the pressure field inside the cavity were plotted against time, for Ψ values from 0.1 – 0.9. A selection of these contour plots can be seen in Figure 9. For all Ψ values, both chambers inside the cavity resonate in unison at the first peak frequency, and therefore the whole cavity is responsible for transmission loss at the first peak frequency.

The second peak frequency is dominated by only one of the two chambers. Cavity 1 is the resonating cavity up until Ψ = 0.4. Ψ = 0.5 and Ψ = 0.6 is where a transition towards Cavity 2 occurs with it being the resonating cavity from Ψ = 0.7 onwards. These two Ψ values are interesting as both chambers resonate at the second peak frequency, similar to both chambers resonating for the first peak frequency. This may be a possible reason for the minimum in Figure 8. where the difference between the first and second peak frequencies is the least, thereby tending towards an increased bandwidth of sound absorption.

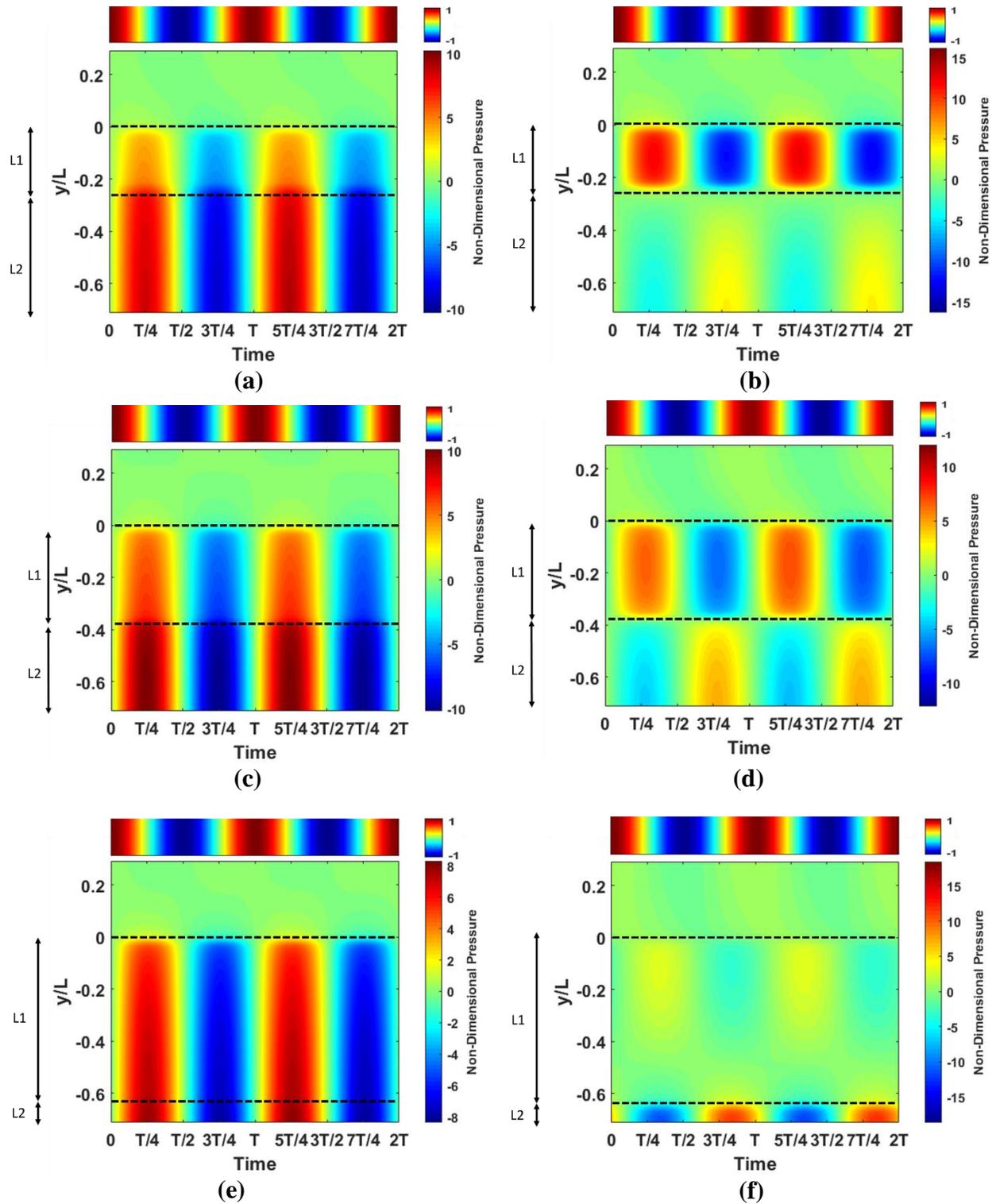

**Figure 9. Contour plots showing change in total acoustic pressure inside the double degree of freedom liner cavity at a transmission loss peak frequency with respect to time period. (a) Ψ = 0.3 at the first peak frequency; (b) Ψ = 0.3 at the second peak frequency; (c) Ψ = 0.6 at the first peak frequency; (d) Ψ = 0.6 at the second peak frequency; (e) Ψ = 0.8 at the first peak frequency; (f) Ψ = 0.8 at the second peak frequency**

## VI. Conclusion

Sensitivity of resonance frequencies of a 2DoF helmholtz resonator compared to a single degree of freedom resonator is studied. It is found that changing the 2DoF cavity volume ratio does not substantially affect the first peak of the resonance frequency however, the difference between the first and second peak reduces with increasing volume ratio and reaches a minimum after which the difference increases again.

Contour plots of the acoustic pressure inside the resonator cavity were used to understand this convergent-divergent behaviour of the second peak frequency. It has been found that the first peak frequency is governed by the whole chamber resonating whereas the second peak frequency is governed by either the top or the bottom cavity resonating depending on the length ratio, $\Psi$. A $\Psi$ of 0.5 – 0.6 is where the resonating cavity starts transitioning from Cavity 1 towards Cavity 2 and at $\Psi = 0.6$, both cavities resonate for the second peak frequency. This may be a possible reason for the minimum in Figure 8. where the difference between the first and second peak frequencies is the least, thereby tending towards an increased bandwidth of sound absorption.

## VII. Future Work

For future studies, the authors will commit to further analysis of experimental data that has already been collected, which will include:
1. Studying the reason behind an increase in the bandwidth of sound absorption on increasing the total number of cavities within the liner body.
2. A study on the positioning of cavities within the whole liner body, to compare against cavities being placed next to each other.
3. Modifying the double degree of freedom resonator geometry based on existing experimental and numerical results to further improve the bandwidth of sound absorption.